\documentclass[pre,amsmath,amssymb,twocolumn,nofootinbib,floatfix]{revtex4-2}
\usepackage[colorlinks=true,linkcolor=blue,urlcolor=blue,citecolor=blue]{hyperref}
\usepackage{graphicx}
\usepackage{bm,color}
\usepackage{times}
\usepackage{color}

\renewcommand{\doi}[2]{\href{http://dx.doi.org/#1}{#2}}
\newcommand{\arxiv}[1]{\href{http://arxiv.org/abs/#1}{#1}}
\newcommand{\link}[2]{\href{http://#1}{#2}}

\newcommand{\Eq}[1]{Eq.~(\ref{#1})}
\newcommand{\Eqs}[1]{Eqs.~(\ref{#1})}
\newcommand{\eq}[1]{(\ref{#1})}
\newcommand{\bea}{\begin{eqnarray}}
\newcommand{\eea}{\end{eqnarray}}
\newcommand{\rme}{\mathrm{e}}
\newcommand{\rmd}{\mathrm{d}}
\newcommand{\nn}{\nonumber}
\renewcommand{\epsilon}{\varepsilon}
\newcommand{\ca}[1]{{\cal #1}}
\newcommand{\be}{\begin{equation}}
\newcommand{\ee}{\end{equation}}
\newcommand{\Fig}[1]{\includegraphics[width=\columnwidth]{./#1}} 

\graphicspath{{figures/},{}}

\renewcommand{\log}{\ln}

\begin{document}

\bibliographystyle{KAY-hyper}

\title{Force Correlator for Driven Disordered  Systems at Finite Temperature}
\author{Cathelijne ter Burg and Kay J\"org Wiese}
  \affiliation{\mbox{Laboratoire de Physique de l’E\'cole Normale Sup\'erieure, ENS, Universit\'e PSL, CNRS, Sorbonne Universit\'e,} \mbox{Universit\'e Paris-Diderot, Sorbonne Paris Cit\'e, 24 rue Lhomond, 75005 Paris, France.}}

\begin{abstract}
When driving a disordered elastic manifold through quenched disorder, the pinning forces exerted on the center of mass are fluctuating,  with mean $f_c=-\overline{F_w} $ and variance $\Delta(w)=\overline{F_w F_0}^c$, where $w$ is the externally imposed control parameter for the preferred position of the center of mass.
$\Delta(w)$ was obtained via the functional renormalization group in the limit of vanishing temperature $T\to 0$, and vanishing driving velocity $v\to 0$. 
There are two fixed points, and deformations thereof, which are well understood: The depinning fixed point ($T\to 0$ before $v\to 0$) rounded at $v>0$, and the zero-temperature equilibrium fixed point ($v\to 0$ before $T\to 0$) rounded at $T>0$. 
Here we consider the whole parameter space of driving velocity $v>0$ and temperature $T>0$, and quantify numerically  the crossover between these two  fixed points. 

\smallskip
\end{abstract} 

\maketitle

\newpage

\section{Introduction}
\subsection{Generalities}

Elastic manifolds driven in a disordered medium  have a depinning transition at zero temperature.  Typical examples are the motion of domain walls in magnets \cite{Barkhausen1919,DurinBohnCorreaSommerDoussalWiese2016,CizeauZapperiDurinStanley1997,terBurgBohnDurinSommerWiese2021}, contact line depinning   \cite{LeDoussalWieseMoulinetRolley2009}, earthquakes  \cite{GutenbergRichter1944,GutenbergRichter1956} and the peeling of a RNA-DNA helix  \cite{WieseBercyMelkonyanBizebard2020}. What these systems have in common is that 
they are governed by an over-damped equation of motion for the interface $u(x, t)$ which is driven through a quenched disordered medium,
\bea
\partial_t u(x,t) & = & \text{ } \nabla^2 u(x,t) + m^2 [w - u(x,t)]  \label{EOM-interface} \\  
&  &   + F\big(x,u(x,t)\big) + \eta(x,t) , \notag\\
w & = & \text{ } v t, \qquad v\ge 0. \notag 
\eea
The disorder forces $F(x,u)$ are short-range correlated, quenched random variables, whereas $\eta(x,t)$ is a thermal noise. Their correlations are 
\bea\label{2}
\overline{F(x,u) F(x',u') } & = & \text{ } \delta(x-x') \Delta_0(u-u') , \\
 \left<\eta(x,t) \eta(x',t') \right> & = & \text{ } 2 T \delta(x-x') \delta(t-t') . 
 \label{3}
\eea
The equation of motion \eqref{EOM-interface} can be studied via field theory. Its principle object is the \emph{renormalised force correlator} $\Delta(w)$. Interestingly, $\Delta(w)$ is the zero-velocity limit of the connected correlation function of the forces acting on the center of mass $u_w = \frac{1}{L^d} \int_x u(x, t) $ \cite{LeDoussalWiese2006a}
\bea
\Delta(w)  & =  & \lim_{v \to 0} \Delta_v(w),  \notag  \\ 
& = &   \lim_{v \to 0} L^{d}m^4\langle [u_w - w][u_{w^\prime} - w^\prime] \rangle^{\rm c} . \label{DeltaDefIntro}
\eea
The functional renormalization group (FRG)   predicts   two distinct universality classes, 
termed {\em depinning} and {\em equilibrium}. 
Equilibrium is the limit of first $v \to 0$ and then $T \to 0$, whereas depinning is the limit of first $T \to 0$ and then $v \to 0$.
In both classes, $\Delta(w)$ has a cusp, and  admits a scaling form 
\be\label{5}
\Delta(w) = m^4 \rho_m^2 \tilde{\Delta}(w/\rho_m). 
\ee 
The characteristic sale $\rho_m$ scales with $m$, 
\be
\rho_m\sim m ^{-\zeta},
\ee
defining a roughness exponent $\zeta$, distinct  between depinning and equilibrium. A second difference is in the  shape  of $\tilde{\Delta}(w)$.

The function $\Delta(w)$ was measured in numerical simulations \cite{
MiddletonLeDoussalWiese2006,RossoLeDoussalWiese2006a}
, and experiments \cite{LeDoussalWieseMoulinetRolley2009,DurinBohnCorreaSommerDoussalWiese2016,terBurgBohnDurinSommerWiese2021,WieseBercyMelkonyanBizebard2019,terBurgRissonePastoRitortWiese2022}. These measurements, both in simulations and experiments, are done by moving the center of the confining potential of strength $m^2$ at a small driving velocity $v$. For {\em depinning}, experiments were performed in soft ferro magnets, both  with SR and LR elasticity \cite{terBurgBohnDurinSommerWiese2021} and DNA/RNA peeling \cite{WieseBercyMelkonyanBizebard2020}.
An experiment in equilibrium is DNA unzipping \cite{terBurgRissonePastoRitortWiese2022}.
 In all cases,  the measured force  correlator $\Delta(w)$ agrees with the predictions from field theory and exactly solved models. It is rounded at a finite driving velocity. While the experiments above are for zero-temperature depinning,  in general the finite driving velocity is not the only perturbation taking us away from the critical point \eq{5}: thermal noise at temperature $T>0$   in Eq.~\eqref{EOM-interface} has to be dealt with.
Apart from the two fixed points depinning and equilibrium, also small deformations of these fixed points are well understood: For depinning, driving at a finite velocity can be accounted for by twice convoluting the zero-velocity fixed point with the response function, which leads to a rounding of the cuspy fixed point \cite{terBurgWiese2020}. On the other hand, the equilibrium fixed point is rounded by a finite temperature, described by a boundary layer \cite{BalentsLeDoussal2004,Wiese2021}. 
%
%
%
 The goal of this paper is to  describe the    crossover  between these two limiting cases. 
We do this by means of numerical simulations. 
At fixed $m^2$, our results are parametrised by $v$ and $ T$. 
 
\subsection{Mean-field description }
Since these questions are  difficult to treat numerically for an interface, our study is done for a single degree of freedom which can  itself be interpreted as the center-of-mass of the interface, or the {\em mean field}. Denoting the center of mass of the interface by $u(t)$, the equation of motion \eq{EOM-interface} and noise correlations \eq{2}-\eq{3} reduce to
\bea
\partial_t u(t) &=&   m^2 [w - u(t)] + F(u) + \eta(t), \label{EOM}\\
 \left<\eta(t) \eta(t') \right> &=& 2 T \delta(t-t') , \\
 \overline{F(u) F(u') } & = &  \Delta_0(u-u') .
\eea
The first term is the force exerted by a confining well, which gets replaced by a Hookean spring with spring constant $m^2$. $F(u)$ is the random pinning force, possibly the derivative of a potential, $F(u) = -  \partial_u V(u)$. Specifying the correlations of $F(u)$ defines the system. Following \cite{terBurgWiese2020}  we  consider forces $F(u) $ that describe an Ornstein-Uhlenbeck (OU) process driven by a Gaussian white noise $\xi(u)$
\bea
\partial_u F(u) &=& - F(u) + \xi(u) \label{FEOM}, 
\\
\left< \xi(u) \xi(u')\right> &=& 2  \delta(u-u'). \notag 
\eea
\begin{figure}[t]
\Fig{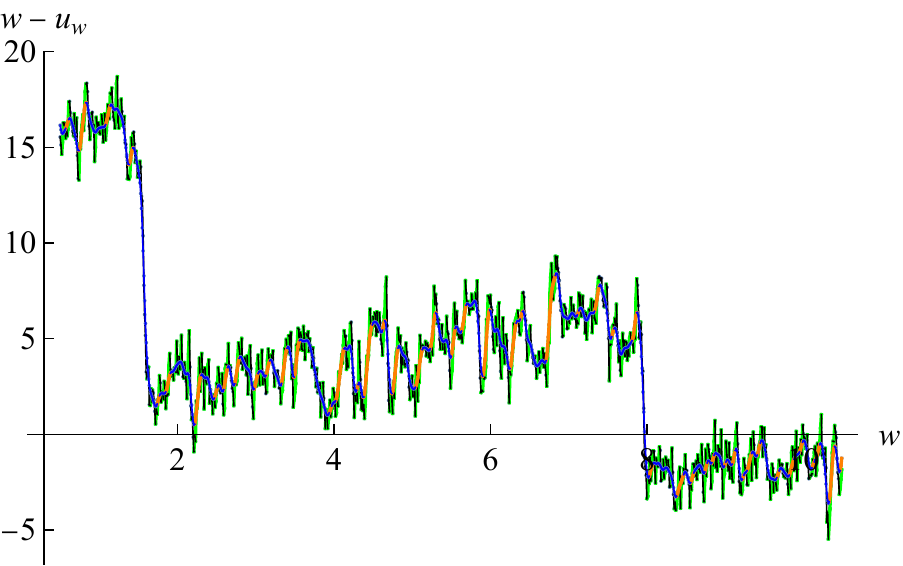}
\caption{$w - u_w$ for $\delta t = 10^{-3}$, $v = 0.01$, $m^2 = 0.1$, $T = 0.5$. (DNS).
In green are the simulation data. The latter are   smoothened to a sixty times larger $\delta t$. This allows us to identify forward moving sections (blue), and backward moving sections (orange).}
\label{f:forceThermal}
\end{figure}%
At small distances $u\ll 1$, the forces $F(u)$ have the statistics of a random walk, thus 
its microscopic limit is the ABBM model \cite{AlessandroBeatriceBertottiMontorsi1990,AlessandroBeatriceBertottiMontorsi1990b}.
At large distances $w\gg 1$ forces are uncorrelated, putting our model in the random-field (RF) universality class. 
%

 Returning to the equation of motion \eqref{EOM}, at zero temperature and at slow driving, most of the time the l.h.s.\ vanishes.  This condition defines the force $F_w$, given $w$, and the associated critical force $f_c$ as 
\bea
F_w & = &   m^2(  u_w-w ) ,  \label{ForceDef} \\ 
f_{\rm c} & := &  \lim_{v\to 0} - \overline{F_w} = \lim_{v\to 0} m^2\overline{(w - u_w ) }.   \label{Fcdef}
\eea
The signs are such that exerting a positive force $f_{\rm c}$ overcomes the pinning forces $F(u_w)$. Due to the thermal noise, $u_w$ can increase even below the threshold force by thermal activation over energy barriers $U$. For sufficiently small velocities, this allows the dynamics to equilibrate with activation times following an Arrhenius law $\tau \sim e^{U/T}$. Thermal fluctuations allow for $u_w$ to go backward, violating the Middleton theorem \cite{Middleton1992} (forward-only motion at $T=0$). 

Fig.\ \ref{f:forceThermal} shows one simulation, with the original trajectory which includes all noise in green. 
Smoothening  it over time allows us to show predominantly  forward movement   in blue and backward movement in orange. 
We see that at this temperature backward movement is substantial. 

The effective disorder is defined as 
\be 
\Delta_{v ,T}(w)  :=   \overline{ F_wF_{w^\prime} }^{\rm c} .   \label{DeltaDef} 
\ee
We have written   subscripts $v, T$ to indicate   that    measurements depend on both $v$ and $T$. Finally, the critical force is related to the area of the  hysteresis loop as 
\be
m^2[\overline{(w - u_w ) }^{\rm forward} - \overline{(w - u_w ) }^{\rm backward}] = 2 f_{\rm c}. 
\ee
Hysteresis is absent in equilibrium where $f = 0$ and maximal at depinning. 
 \begin{figure}[t]
\fboxsep0mm
\setlength{\unitlength}{1cm}
{\begin{picture}(8.6,7.0)
\put(0,0){\includegraphics[width=8.6cm]{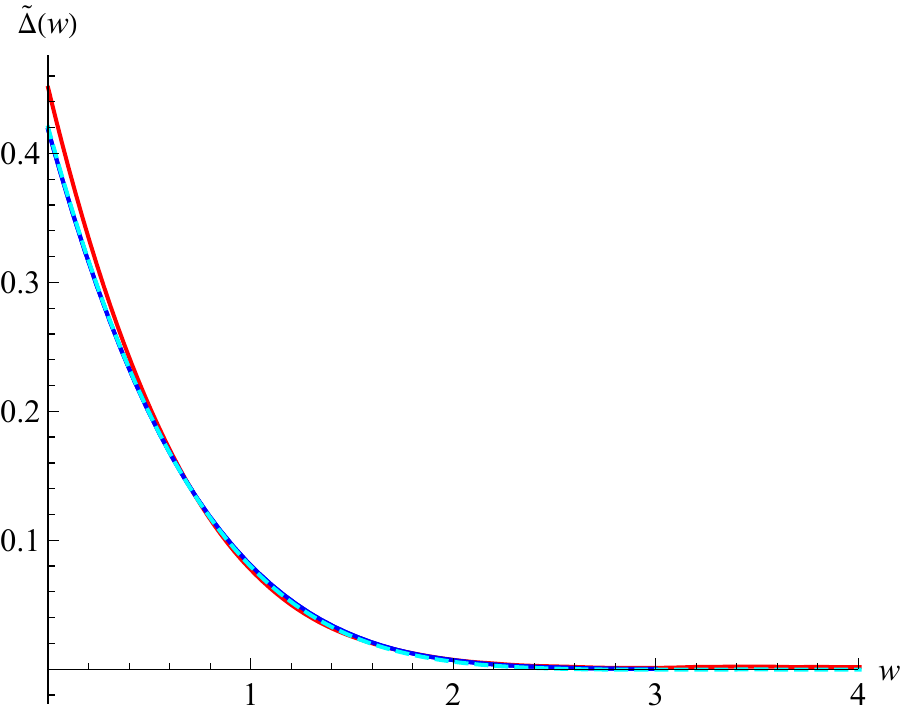}}
\put(2.5,1.9){\includegraphics[width=6.0cm]{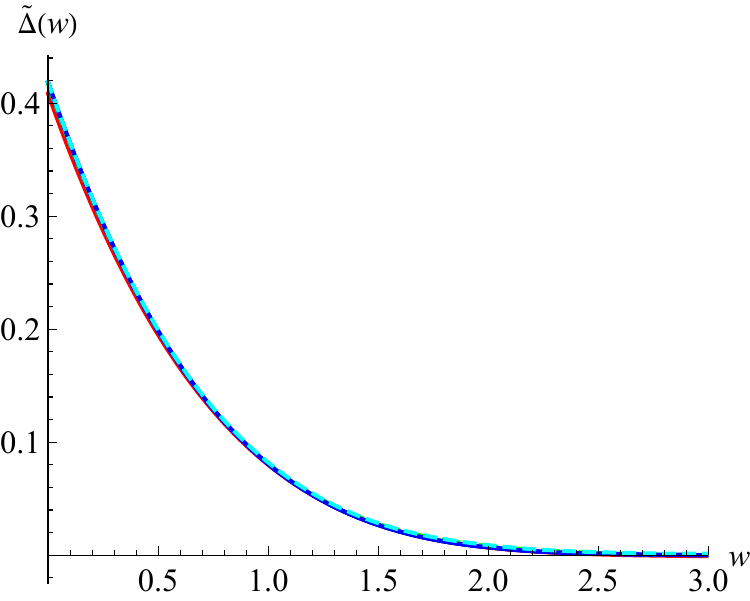}}
\end{picture}}
\caption{$\tilde{\Delta}(w)$ for the Sinai model (blue) obtained by numerical integration of \Eqs{18}-\eq{tDelta-Sinai}. It is compared to the energy minimisation for  $m^2 = 0.1$ (red), $m^2 = 0.01$ (cyan dashed) indistinguishable from the theory. Statistical errors are within the line thickness. Inset: Idem for the OU model in \Eq{FEOM}.} 
\label{f:sinai}
\end{figure}

\subsection{Review of known results}
Before we present our findings for the questions posed in the introduction, let us review what is known for  a single perturbation.

\subsubsection{Equilibrium fixed point}
\label{s:TempEquilibrium}

 \begin{figure}[b]
\fboxsep0mm
\setlength{\unitlength}{1cm}
{\begin{picture}(8.6,7.0)
\put(0,0){\includegraphics[width=8.6cm]{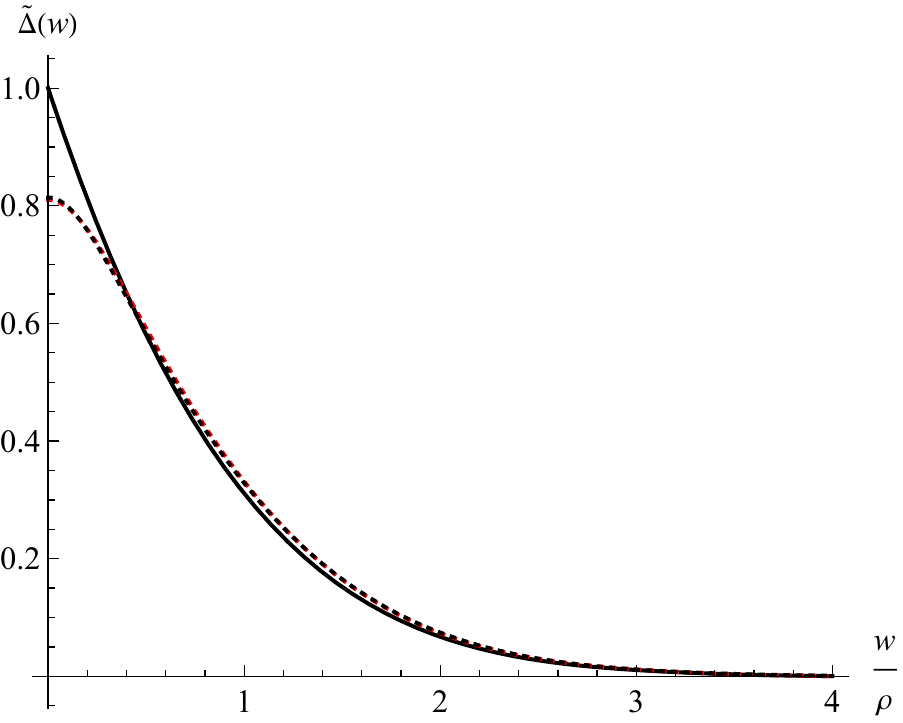}}
\put(2.5,2.4){\includegraphics[width=6.0cm]{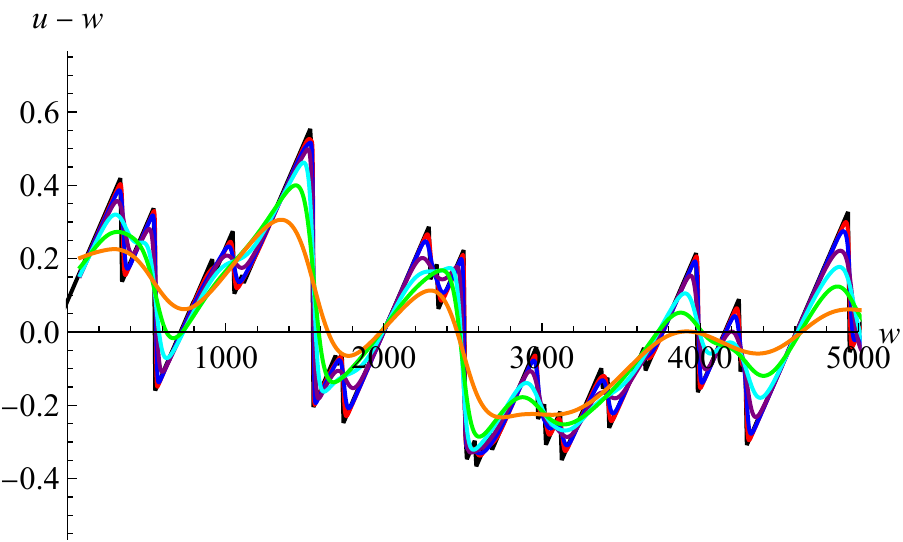}}
\end{picture}}
\caption{Boundary layer analysis for equilibrium random field disorder (EM) for the dimensionless rescaled disorder $\tilde{\Delta}(w)$ and rescaled to have unit amplitude and slope 1 at $w = 0$.  Black solid, $v= 0, T = 0$ fixed point, black dashed, numerical measurement at $m^2 = 0.01$, $T = 2$, red dotted, thermal boundary layer ansatz using equations \eqref{boundaryLayerApprox}. Inset:  The effective force at various $T$.  } 
\label{f:BoudaryLayerForce}
\end{figure}

\begin{figure*}[t]
{\setlength{\unitlength}{1cm}\begin{picture}(17.8,5.9)
\put(0,0){ \includegraphics[width=.33\textwidth]{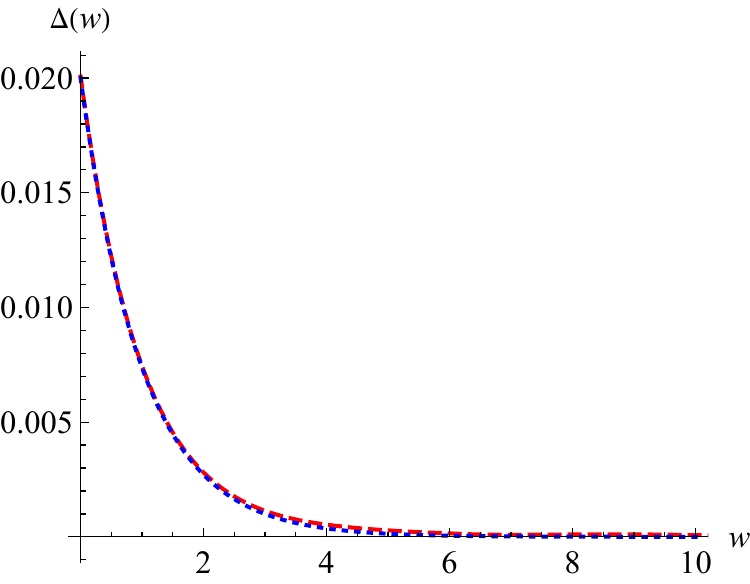}}
\put(5.75,0.0){ \includegraphics[width=.33\textwidth]{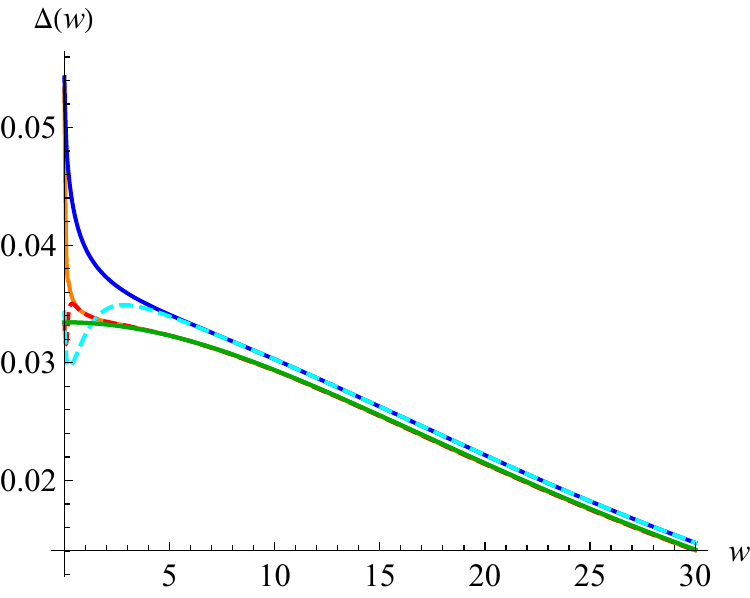}}
\put(7.75,2.0){ \includegraphics[width=.2\textwidth]{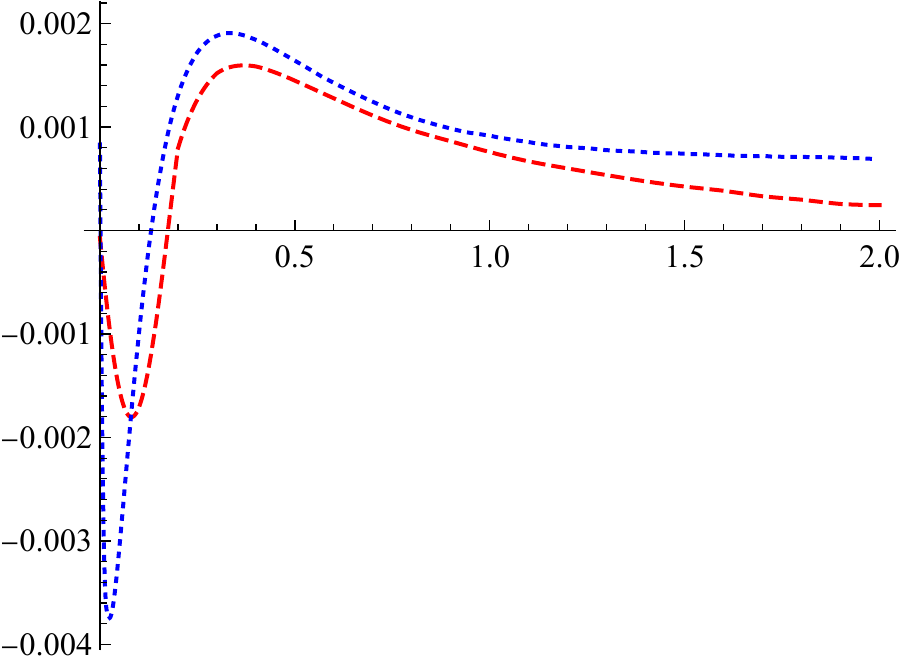}}
\put(11.5,0.0){ \includegraphics[width=.33\textwidth]{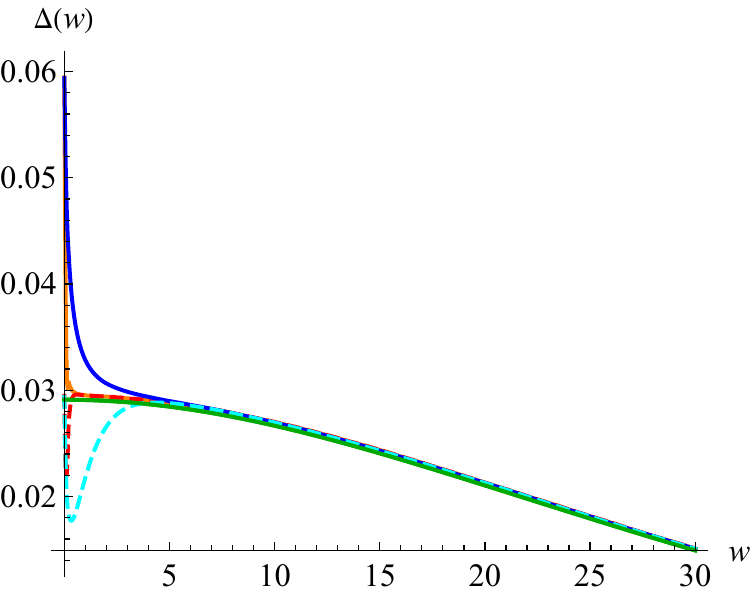}}
\put(13.5,1.8){ \includegraphics[width=.2\textwidth]{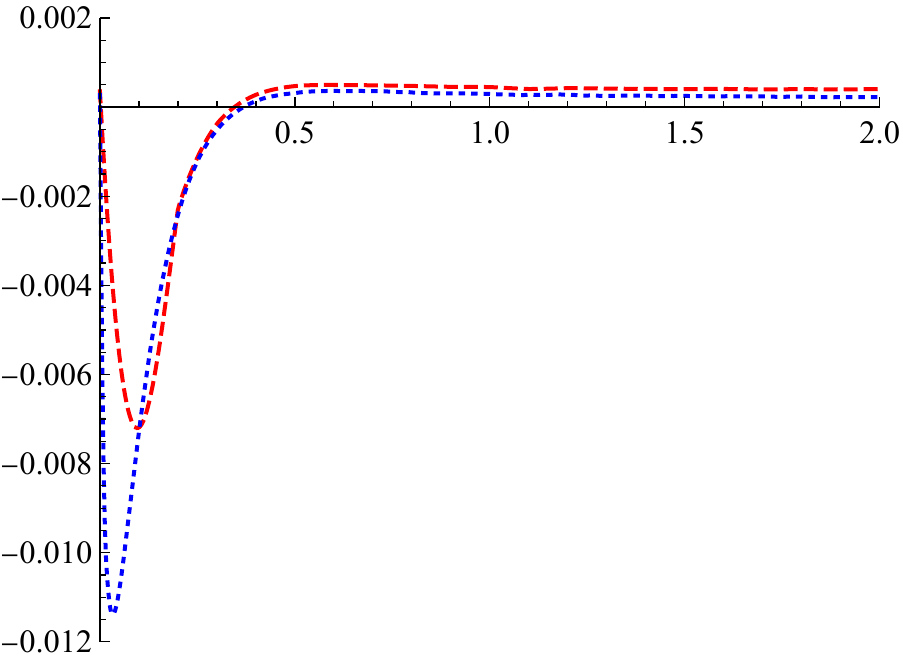}}
\put(3.0 ,5.0){(a)}
\put(9.0 ,5.0){(b)}
\put(15.3 ,5.0){(c)}
\end{picture}}
\caption{(a) Comparison of the prediction \eqref{thermalSubtraction} (blue) for $m^2 = 0.01$, $v = 10^{-3}$, $T = 2$ to a measurement of the noise correlations in the absence of disorder. 
(b) and (c) are in presence of quenched disorder.
\\
Subfigures (b) and (c): For $m^2 = 0.01$, $T = 2$ (b) and $T= 3$ (c) comparison of the equilibrium $\Delta_{T}^{\rm eq}(w)$ (green, EM) to $\Delta_{v, T}(w)$ at $v = 10^{-3}$ (blue solid, DNS) and $v = 10^{-4}$ (orange solid, DNS). In dashed cyan/red, we show the combination \eqref{comb}. This correctly captures the amplitude, but a signal of anti-correlations remains. In the inset we show   the difference $\delta\Delta_{v,T}(w/v) $, which quantifies the corrections due to non-equilibration.}
\label{f:SubtractionT2T3}
\end{figure*}

The   zero-temperature equilibrium fixed point can be measured by energy minimisation (EM)  at fixed $w$ of 
\be 
\mathcal{H}_w(u) = \frac{m^2}{2}(u - w)^2 + V(u) , \label{EnergyMinimisation}
\ee
 see appendix \ref{s:Numerical implementations} for implementation details.
The random potential is given by $V(u)  = - \int F(u) {\rm d}u$.  For the random-field (RF) disorder relevant for \Eq{FEOM}, the  model is  known as the Sinai model,   introduced in \cite{Sinai1983}.  The effective force  correlator   reads (see \cite{LeDoussal2010}, with corrections in \cite{Wiese2021}) 
 \bea\label{18}
\Delta(w) &=& m^4 \rho_m^2 \tilde \Delta(w/\rho_m )  ,\\
\rho_m &=&    2^{\frac 2 3}  m^{-\frac 43}\sigma^{\frac 1 3} ,
\label{rho-m-Sinai}
\eea

\begin{align}
\tilde \Delta(w) \ = &  -\frac{  \rme^{-\frac {w^3}{12}}}{ 4\pi^{\frac 3 2}\sqrt w}\int\limits _{-\infty}^{\infty} \rmd \lambda_1 \int\limits_{-\infty}^{\infty} \rmd \lambda_2 \,\rme^{-\frac{(\lambda_1{-}\lambda_2)^2}{4 w  }} \label{tDelta-Sinai} \\
\ & \times \rme^{ i \frac w {2} (\lambda_1+\lambda_2)} \frac{\mbox{Ai}'(i \lambda_1)}{\mbox{Ai}(i \lambda_1)^2}
\frac{\mbox{Ai}'(i \lambda_2)}{\mbox{Ai}(i \lambda_2)^2} \notag \\
\ &\times \!\bigg[1{+} 2 w\frac{\int_0^\infty \rmd V \rme^{w V}  \mbox{Ai}(i \lambda_1{+} V) \mbox{Ai}(i \lambda_2{+} V) }{\mbox{Ai}(i \lambda_1) \mbox{Ai}(i \lambda_2)} \bigg]\!.\notag 
\end{align}
The roughness exponent is identified from \Eq{rho-m-Sinai} as $\zeta = 4/3$.  Fig.\ \ref{f:sinai} shows 
  in blue   the analytical solution  of \Eqs{18}-\eq{tDelta-Sinai}. 
 In red and  cyan are numerical simulations of \Eq{EnergyMinimisation} for uncorrelated forces, constant in an interval of size one, and unit variance, i.e. $ \langle V(u) - V(u^\prime) \rangle \simeq |u - u^\prime| $. Already for $m^2 = 0.01$, the simulation has converged to the theory. The inset shows comparison to the model  of OU forces defined in \eq{FEOM}, which belongs to the same universality class.  
 
 At a finite temperature, thermal fluctuations smoothen the shocks and round the cusp in a boundary layer $u \sim T$.  This {\em thermal rounding} is shown in Fig.~\ref{f:BoudaryLayerForce}.  The size of the boundary layer can be estimated from the FRG \cite{Wiese2021} (see appendix \ref{A:Boundary layer}) 
\bea
\Delta^{\rm eq}_T(w)  & = & \text{ } \mathcal{A}_{T} \Delta^{\rm eq} ( \tilde{w} ),  \label{boundaryLayerApprox}  \\ 
\tilde{w} & = &  \text{ }  \sqrt{w^2  + t^2} , \qquad \frac{t}{\rho_m} = \frac{3}{\epsilon}  \frac{2T m^2}{\Delta(0)} ,  \label{boundaryLayerApprox2} \qquad  \\
\mathcal{A}_{T} & = & \text{ }  \frac{\int_{0}^{\infty} {\rm d}w \Delta^{\rm eq} ( w )}{\int_{0}^{\infty} {\rm d}w \Delta^{\rm eq} ( \tilde{w} )}  . 
\label{boundaryLayerApprox3}
\eea
The amplitude $\mathcal{A}_{T} $ ensures normalisation, i.e.\ that the area under $\Delta(w)$ is preserved in the presence of  thermal rounding. 
Since from \eqref{boundaryLayerApprox2} one sees that $w \sim t \sim \rho_m$, the RHS of  \eqref{boundaryLayerApprox2} is dimensionless. This defines the dimensionless temperature $T_m  \sim T m^{\theta}$, scaling with its own exponent  
\be
\theta = d - 2 + 2 \zeta.
\ee $\theta$ is called the \emph{equilibrium energy exponent} 
As we show in appendix \ref{A:Boundary layer},
an alternative expression for the boundary layer  is given by 
\bea\label{DiffKernelsupp} 
\Delta_{T}(w)  &=&    \int_{-\infty}^{\infty} \rmd u\,  \Delta(u) G(u-w,\tau), \\
G(u,\tau)  &=& \frac{1}{ \sqrt{4\pi \tau}} \rme^{-\frac{u^2}{4 \tau }} \\
\tau &=& \frac{t^2}{\pi} - \frac{2(\pi - 2)t^3}{\rho_m \pi^2} +  \mathcal{O}(t^4).
\eea
where $G(u, \tau)$ is a diffusion kernel. A  delicate question is what the dynamical exponent $z$ is in equilibrium. The observation that $z = 2$ in both the free theory as well as at depinning suggests that this likely holds also in equilibrium. Finally, the pinning force $f_{\rm c} =  0$ in   equilibrium. 

\subsubsection{Depinning fixed point}

For depinning the effective disorder \eqref{DeltaDefIntro} is given by  \cite{LeDoussalWiese2008a,terBurgWiese2020}
\bea
\Delta(w) &= &  m^4 \rho_m^2 \tilde{\Delta}_{\rm Gumbel} (w/\rho_m) , \label{DeltaGumbel}\\
\tilde{\Delta}_{\rm Gumbel} (w)& = & \frac{w^2}{2} + {\rm Li }_2(1 - e^{|w|}) + \frac{\pi^2}{6},   \label{DeltaGumbel2}  \\
\rho_m & = &  \frac{1}{2m^2 \log{(m^{-2})}}.       \label{DeltaGumbel3}
\eea
The roughness exponent   is $\zeta = 2^-$; the dynamical exponent is $z = 2^-$ \cite{terBurgWiese2020}. In the simulations, we can measure \eqref{DeltaGumbel} at zero velocity, by  moving the parabola from $w \to w + \delta w$ and waiting for the dynamics to cede. In an experiment, performed at finite $v$, $\Delta(w)$ is rounded by the driving velocity \cite{terBurgWiese2020}
\be
\Delta_v(w- w^\prime) = \iint R(t)R(t^\prime) \Delta( w- w^\prime - v(t - t^\prime)).  \label{VelRounding}
\ee 
By construction,  $\int_t R(t) = 1$ and the integral of  $\Delta_v(w)$  is independent of $v$. At small $v$, \Eq{VelRounding} can be approximated by 
\be\label{vel-round-approx}
\Delta_v(w) = \frac1{\ca N}\Delta(\sqrt{w^2+(v\tau)^2}),
\ee
where $\ca N$ is chosen s.t.~$\int_w\Delta_v(w) =\int_w \Delta(w) $. 

For $v=0$ the critical force $f_{\rm c}$ is defined in \Eq{Fcdef}. For $v>0$, the combination  $m^2  \overline{(w - u_w ) }$  increases  to \cite{terBurgWiese2020} 
\be 
 m^2  \overline{(w - u_w ) } \approx f_{\rm c}\Big|_{v=0} + \eta v + \ca O(v^2).   \label{FcdefFiniteV}
\ee
Here $\eta$ is the viscosity, set to $\eta=1$ in \Eq{EOM-interface}.

 \section{Results in the general situation}
   \label{s:Delta}

We now present our numerical results, mostly obtained by direct numerical simulation (DNS). First in  section \ref{Equilibrium regime and thermal peak} we  check  
\Eqs{18}-\eq{boundaryLayerApprox3} for equilibrium. 
In section \ref{Order parameters} we discuss several order parameters characterizing the crossover between equilibrium and depinning. 
Section \ref{Scaling collapses} shows that with the rescalings established so far, we can collapse all our data.

\subsection{Thermal peak in the equilibrium regime}\label{Equilibrium regime and thermal peak}
\begin{figure}[t]   
   \includegraphics[width=\columnwidth]{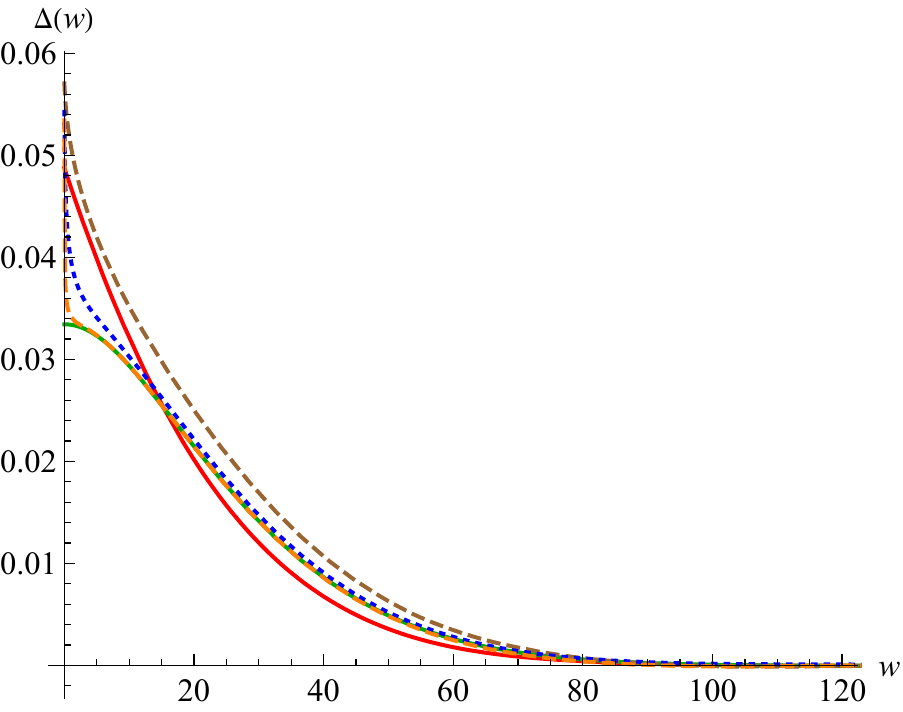}
      \caption{ The equilibrium regime for $T= 2$ with the zero temperature fixed point in red and $\Delta_T(w)$ (EM) shown in green. Simulation of \eqref{EOM} shows $v = 0.01$ (brown), $v = 10^{-3}$ (blue) and $10^{-4}$ (orange) we show $\Delta_{v, T}(w)$ (DNS). For the smallest two driving velocities the agreement is excellent, and the thermal peak, rounded by the driving velocity,  is clearly visible. The largest velocity no longer obeys the decomposition \eqref{EqCond} and belongs to the crossover regime. }
              \label{f:PertEquil}
\end{figure}

In Fig.~\ref{f:PertEquil} we show the results of numerical simulations of $\Delta_{v, T}(w)$ in the near-equilibrium regime. The presence of the thermal noise leads to a thermal peak (TP) at small $w$. 
In absence of disorder it reads 
\bea
\Delta^{\rm {\rm TP}}_v(w-w')  & = &    2Tm^4 \int_{-\infty}^{\infty} R (t, \tau) R (t^{\prime} , \tau) \, \rmd\tau   \nn  \\ 
& =  &   Tm^2e^{-m^2 |t - t^\prime|}   \nn \\
& =  &   Tm^2e^{-m^2 |w - w^\prime|/v} .  \label{thermalSubtraction}
\eea 
Here $R (t) = \Theta(t)e^{-m^2 t} $ is the response function of the free theory. This is checked in Fig.\ \ref{f:SubtractionT2T3}(a). 

Let us now turn back to the disordered case, at finite velocity $v>0$ and finite temperature $T>0$. We make the ansatz 
\be 
\Delta_{v ,T}(w) =    \Delta^{\rm eq }_{T}(w) + \Delta^{\rm {\rm TP}}_v(w) + \delta\Delta_{v,T}(w) .   \label{deltaVcontr} 
\ee
The first term is the relevant result for $v=0$. The second term is the contribution \eq{thermalSubtraction} from the thermal noise. 
If the driving velocity is small enough for the dynamics to equilibrate, then we expect the third term $ \delta\Delta_{v,T}(t)$ to vanish, or at least to be small. 
%
%
%

Figs.~\ref{f:SubtractionT2T3}(b)-(c) show the combination 
\be
\label{comb}
\Delta_{v ,T}(w) - \Delta^{\rm {\rm TP}}_v(w) =  \Delta^{\rm eq }_{T}(w)  + \delta\Delta_{v,T}(w) , 
\ee
for $T = 2$ (b) and $T = 3$ (c). While $\Delta^{\rm {\rm TP}}_v(w)$  correctly subtracts the thermal noise at $w=0$, the remaining term $ \delta\Delta_{v,T}(t) $ is visible.
In the inset, we show  $ \delta\Delta_{v,T}(t) $, i.e.\ the error we make in the approximation $\Delta_{v ,T}(w) \approx   \Delta^{\rm eq }_{T}(w) + \Delta^{\rm {\rm TP}}_v(w) $. We see that despite a difference of $v$ by a factor of ten, the rescaled combination $\delta\Delta_{v,T}(t=w/v) $ at small $t$ depends little on $v$. 
%
%
This estimates the boundary layer in our example to be  $\delta t \approx 2$. 
\begin{figure}[t]
\fboxsep0mm
\setlength{\unitlength}{1cm}
{\begin{picture}(8.6,6.9)
\put(0,0){\includegraphics[width=8.6cm]{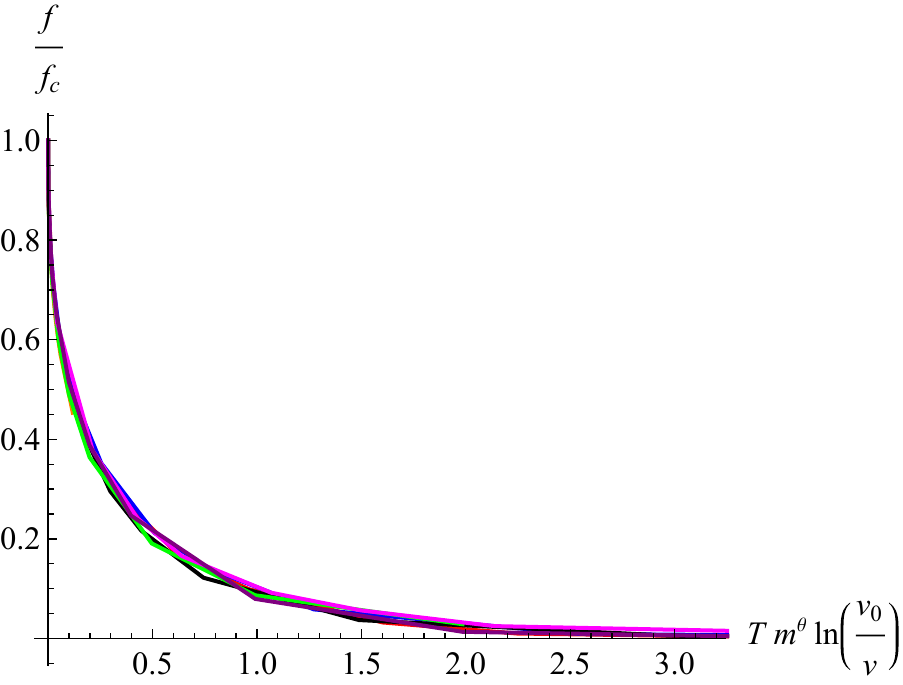}}
\put(2.3,1.2){\includegraphics[width=6.2cm]{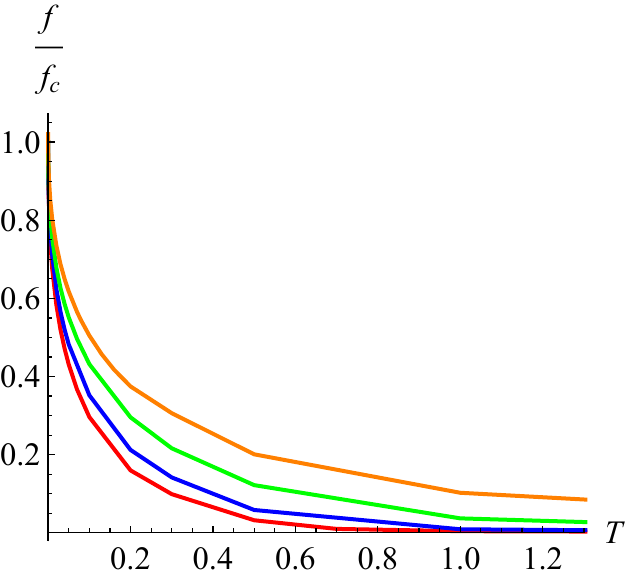}}
\end{picture}}
\caption{Scaling collapse of the measured force  for   $T>0$, and different $m$, $v$. We found an optimal collapse for $v_0 = 1$, but any $v_0$ of the same order of magnitude does   well.} 
\label{f:ForceCollapse}
\end{figure}%
\begin{figure}
\Fig{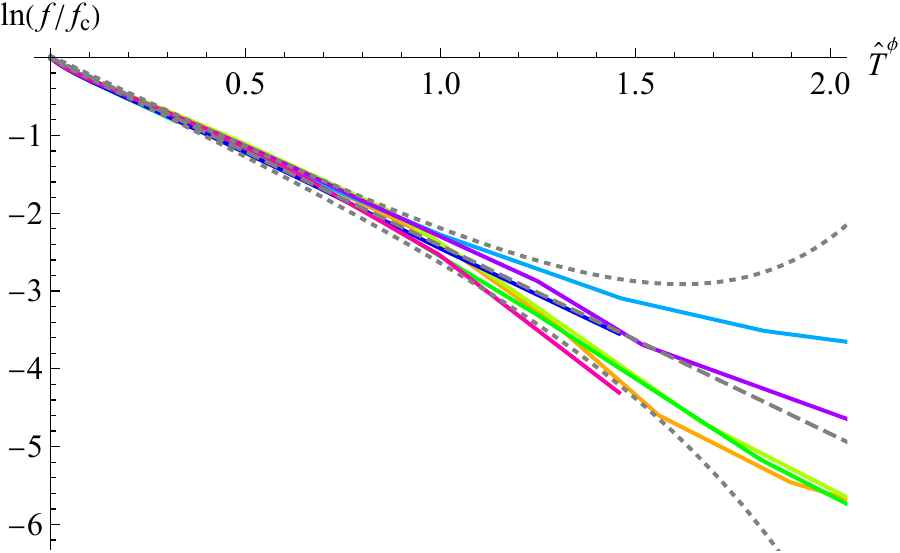}
\caption{This plot shows that $f/f_c$ from Fig.~\ref{f:ForceCollapse} is a stretched exponential, with an exponent of about 0.55. Gray dashed is a fit to a linear function, gray dotted putative error bars.}
\label{f:f-collapse-log}
\end{figure}

\Eq{deltaVcontr} approximately predicts the amplitude for equilibrium as
 \be
 \Delta_{v, T}(0) \approx \Delta_T(0)+  m^2 T.  \label{EqCond}
 \ee
As can be seen in  Fig.~\ref{f:PertEquil}, this relation breaks down for $v = 0.01$, corresponding to $\hat T= Tm^{2/3}\log{(v)} \approx 2 $. A look at  Fig.~\ref{f:ForceCollapse}, discussed next,  shows that there $f/f_{\rm c} \approx 0.05$, which signals the  approach to the crossover regime.

 \subsection{Order parameters}
 \label{Order parameters}
 \subsubsection{The mean force as an order parameter}
\label{s:ForceOrderParameter}
 
\noindent The measured pinning  force 
\be
f:= m^2\overline{(w-u_w)}  - v, 
\ee 
  is maximal for depinning at temperature zero, and vanishes in equilibrium. It 
 is a natural candidate for an order parameter. 
We define \be
\Psi_f:= \frac f{f_{\rm c}} , 
 \ee 
 which vanishes in equilibrium and is $1$ at depinning. 
The inset of  Fig.~\ref{f:ForceCollapse} shows  this   force ratio for different $m^2, T$ and $v$, for $v = 10^{-2}, 10^{-3},10^{-4}$, $m^2 = 0.1-10^{-3}$ and $T =\in [ 0,2]$.  Using that the dimensionless temperature is   $T m^{\theta}$ and velocity and temperature are related by Arrhenius' law as $\ln({v}) \sim 1/T $, a natural ansatz for a scaling parameter is 
\be\label{hatT}
\hat T:= Tm^{2/3}\ln(1/v). 
\ee
This collapses all curves on a single master curve, as shown in the  main plot of Fig.~\ref{f:ForceCollapse}.    We can go one step further. To do so, let us plot the log of $f/f_{\rm c}$ as a function of $\hat T^{\phi}$. We find on Fig. \ref{f:f-collapse-log}
an almost linear behaviour for an exponent $\phi=0.55$, with slope $-2.41$. Thus  
\be
\frac f{f_{\rm c}} \approx \rme^{- \left(\frac{\hat T}{\hat T_c}\right)^\phi}, \qquad \phi\approx 0.55, \quad \hat T_c \approx 0.2. 
\ee
is a stretched exponential. Note that if the fit is attempted close to $f\approx f_c$, one can also conclude on $\phi\approx 0.51$. If we restrict to 10 percent deviation, this allows for $\phi$ in the range $\phi \in [0.51,0.56]$.  We   expect the regime  $f/f_{\rm c}\to 1$ to be governed by the depinning fixed point, and  $f/f_{\rm c}\to 0$ by the equilibrium fixed point. The crossover regime should be best  visible for  $f/f_{\rm c}\approx 1/2$ .

\begin{figure}[ht]
\includegraphics[width=9.0cm]{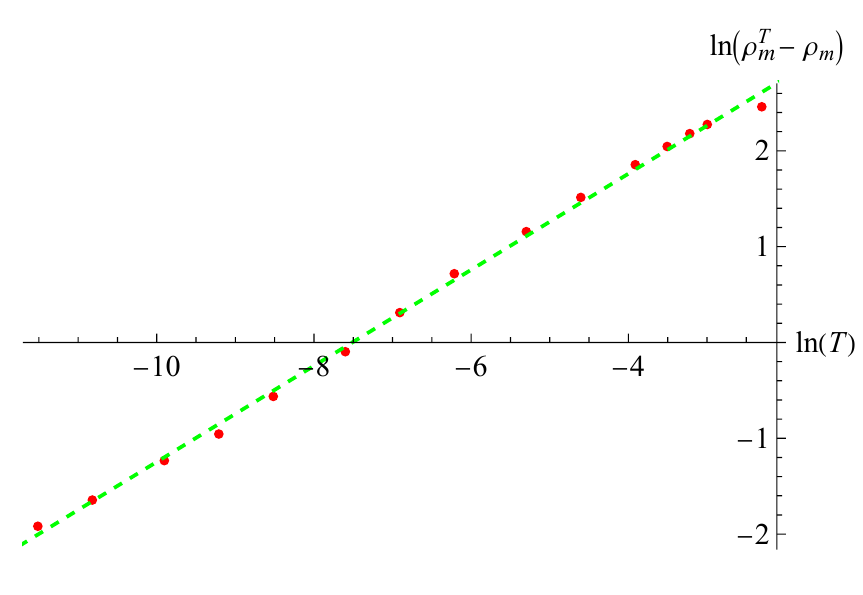}
\caption{Check of the scaling relation  \eqref{DeltaTScaling2}. The green dashed line has a slope of $\phi=0.50$. }
\label{f:lnrhooflnv}
\end{figure}

\subsubsection{The correlation length as an order parameter}
\label{s:The correlation length as an order parameter}

\begin{figure}[tb]\fboxsep0mm%
{\setlength{\unitlength}{1cm}\begin{picture}(8.5,7.1)
\put(0,0){\includegraphics[width=8.5cm]{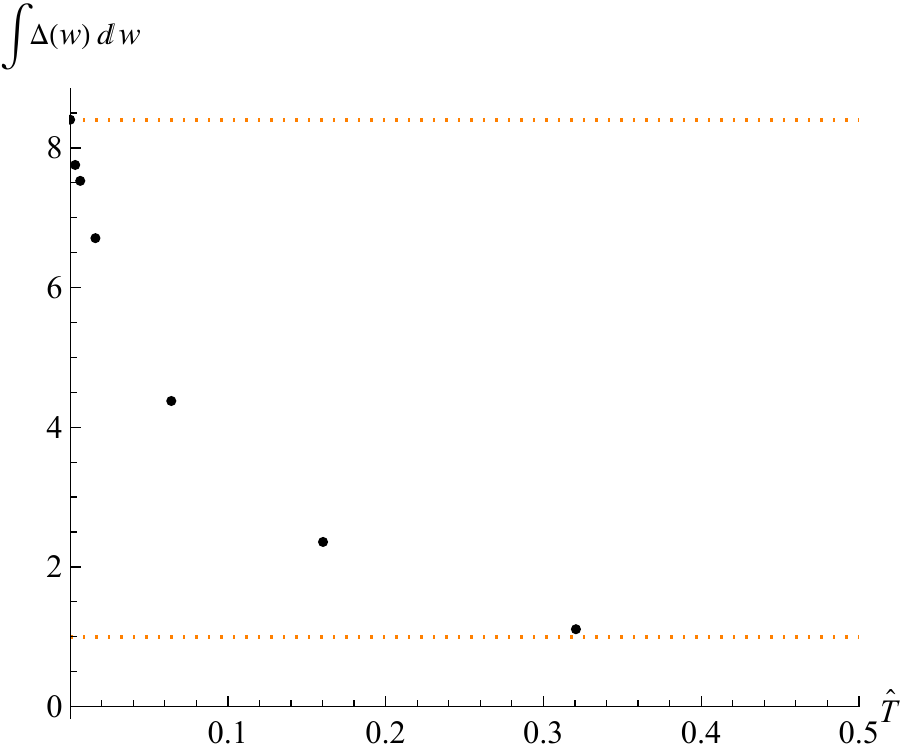}}
\put(2.9,2.3){ \includegraphics[width=0.64\columnwidth]{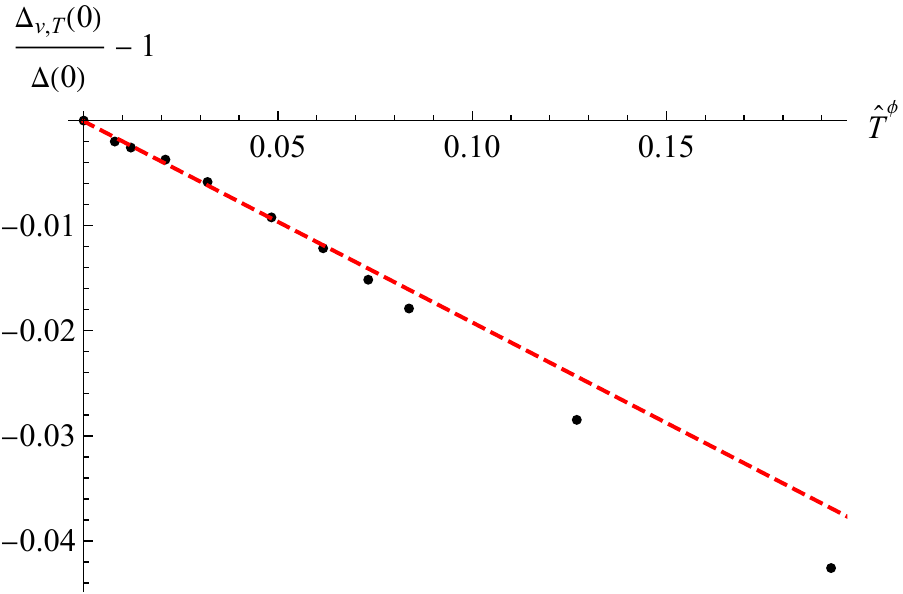}}
\end{picture} 
}  
\caption{The change with temperature of the area (main plot) and amplitude at $w=0$ (inset) for $\phi= 0.56$.  }
\label{f:Area}
\end{figure}

 In section \ref{s:ForceOrderParameter} we established the mean force as an order parameter between equilibrium and depinning. While this is the most robust quantity we found, there are other quantities one might use. The first is the correlation length $\rho_m$, which decreases with temperature compared to its value at depinning. In all of the following, we will denote by $\rho_m$ the $T = 0$ value at depinning, and by $\rho_m^T$ the finite-temperature value.  If one considers  zero-temperature depinning as a reference point, then at small temperatures
 \be
\rho_m - \rho_m^T \sim    \hat{T}^{\phi}  \label{DeltaTScaling2} , 
\ee
with $\phi = 0.50 \pm0.02$, see Fig.~\ref{f:lnrhooflnv}.

\subsubsection{The disorder amplitude as an order parameter}
Assuming that 
\be
\Delta_{v, T}(w)  = m^4 (\rho_m^T)^2 \tilde{\Delta}(w/\rho_m^T), \label{DeltaTScaling1} \\ 
\ee
 the amplitude $\Delta_{v,T}(0)$ at small $T$ should behave as 
\be
\frac{ \Delta_{v, T}(0) }{\Delta(0)}-1 \simeq   \left( \frac{\rho_m^T }{\rho_{\rm m} }\right)^2- 1   \sim      T^{\phi} + \ca O(T^{2\phi}) .
\ee 
Our measurements presented in the inset of Fig.~\ref{f:Area}  are   consistent with an exponent in the range  $\phi \in [0.5, 0.6] $.

\subsubsection{The disorder integral as an order parameter}

Both the correlation length as well as the amplitude are very sensitive to details of the rounding around the cusp. More robust is to consider the area under $\Delta(w)$. The above relations above would imply that 
\be
\frac{\int_{w>0} \Delta_{v, T}(w) }{\int_{w>0} \Delta(w)}\sim \left( \frac{\rho_m^T}{  \rho_m}\right)^{\!3} \sim T^\phi + \ca O(T^{2\phi}). 
\ee
Our data are consistent with an exponent in the range of $\phi=0.5$ to $\phi=0.67$,   favoring the upper end. 
The decrease of the area is shown in Fig.~\ref{f:Area}. The orange dashed lines are references for depinning (top) and equilibrium (bottom).

\subsection{Scaling   close to equilibrium and depinning}
\label{Scaling collapses}

Let us next consider scaling  close to equilibrium  and depinning. 
\subsubsection{Scaling near equilibrium} 
\begin{figure}[t]
{\setlength{\unitlength}{1cm}\begin{picture}(10,6)
\put(0,-0.5){ \includegraphics[width=\columnwidth]{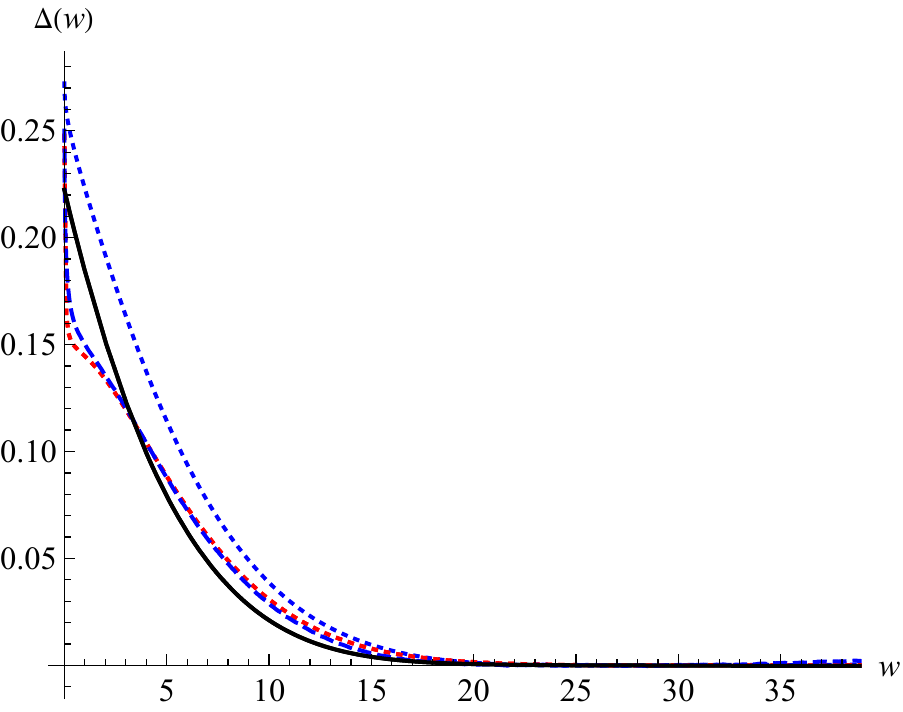}}
\put(2.3,1.3){ \includegraphics[width=0.7\columnwidth]{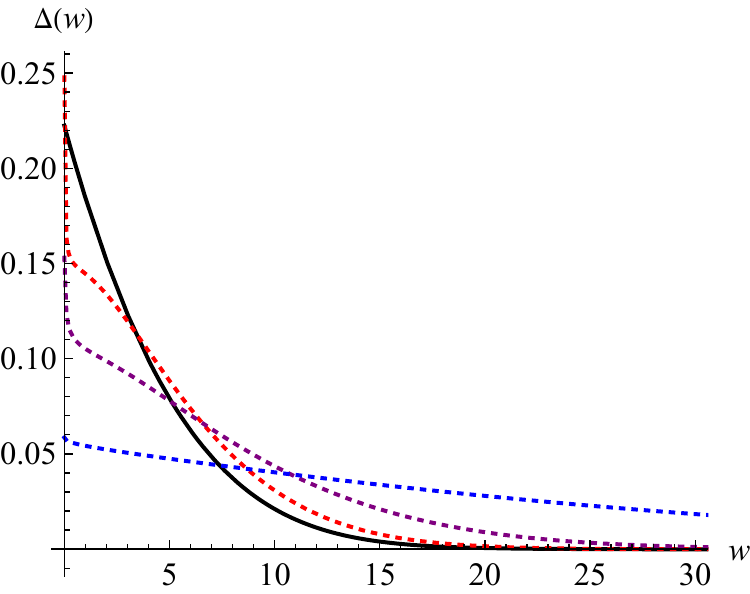}}
\put(1.3,3.3){ \text{I}}
\put(0.7,2.5){ \text{II}}
\end{picture} 
}  
\caption{Scaling of $\Delta(w)$ (DNS). The inset shows the equilibrium fixed point (EM) at $m^2 = 0.1$ (black) and $\Delta_{v, T}(w)$ for $m^2 = 0.1$ (red),  $m^2 = 0.05$ (purple) and $m^2= 0.01$ (blue)  for $T = 1, v = 10^{-3}$. Main plot shows the collapse of $m^2  = 0.01$ onto $m^2 = 0.1$ by I) rescaling only $m^2$, II) rescaling in addition $Tm^{\theta}$.   }
\label{f:MScalingDelta}
\end{figure}

For  equilibrium   
\be
\Delta(w) \sim m^{4/3}\tilde{\Delta}(w m^{4/3}) \label{DeltaMScaling}.
\ee
While this scaling  holds for all $w$ at the zero-temperature fixed point, the scaling within the boundary layer is more subtle. Consider first   the inset of Fig.~\ref{f:MScalingDelta}. In black is shown the equilibrium fixed point for $m^2 = 0.1$; red/blue/purple show from top to bottom $m^2 = 0.1, 0.05, 0.01$ for $T = 1$, $v = 10^{-3}$. In the main plot we perform a scaling collapse from $m^2 = 0.01$ onto $m^2  = 0.1$. In blue dotted (I), we rescaled by only accounting for the difference in mass, i.e. with $m^{4/3}$. One sees it clearly deviates. To obtain a full scaling collapse, one needs to also scale   temperature with its corresponding dimension, i.e. $Tm^{-\theta}$ with $\theta = 2/3$, leading to the blue-dashed curve (II). The remaining offset comes from the velocity which also scales with $m^2$.  Using that $z = 2$ both in the free theory and at depinning, suggest a scaling of $v \sim m^{2/3}$. A look at the size of the boundary layer of the thermal peak suggests that the driving velocity should be reduced by a factor of 2, which is approximately consistent with the above scaling. 

\subsubsection{Scaling near depinning} 

\begin{figure}[tbh]
{\setlength{\unitlength}{1cm}\begin{picture}(10,6)
\put(0,-0.5){ \includegraphics[width=\columnwidth]{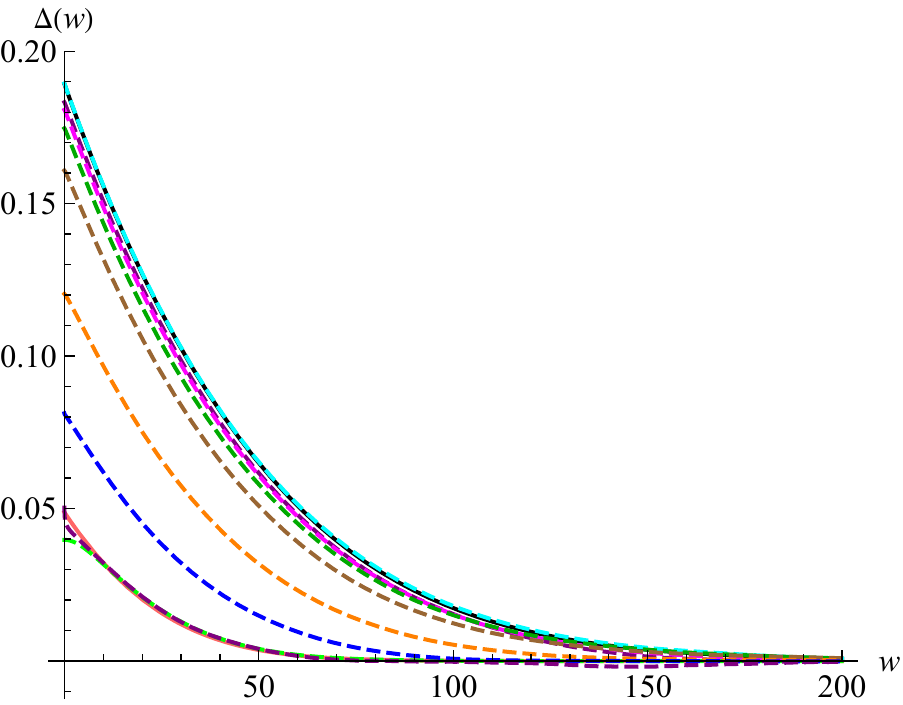}}
\put(2.9,1.5){ \includegraphics[width=0.6\columnwidth]{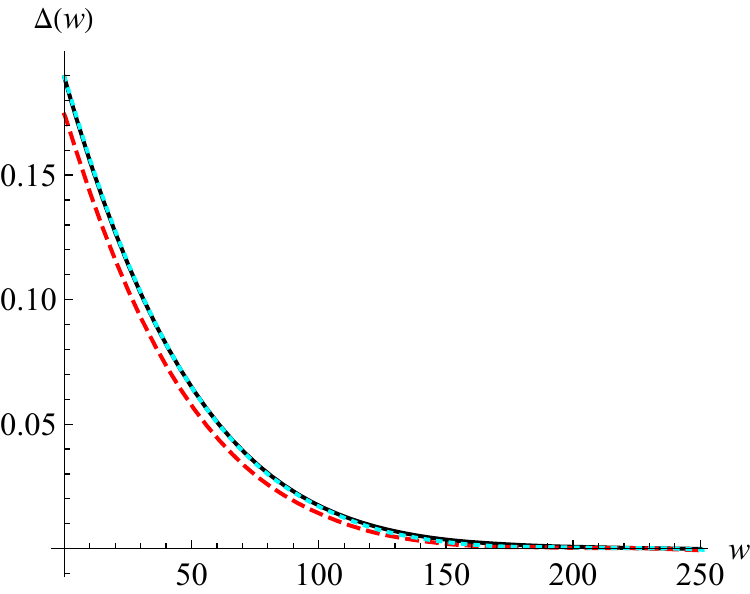}}
\end{picture} 
}  
\caption{$\Delta_{v, T}(w)$ (DNS) for $m^2 = 0.01$, $v= 10^{-3}$ and $T \in [0,0.005, 0.01 ,0.02, 0.05, 0.2, 0.5, 1]$ from equilibrium (red, bottom, EM) to   depinning (black, top). The inset shows scaling collapse using the scaling relation in \eqref{DeltaTScaling1} for $T = 0.02$. $Tm^{\theta} \log{1/v} = 0.03$ indeed close to depinning. Brown curve corresponds to $Tm^{\theta} \log{1/v} = 0.075$ already at 60 $\%$ of the maximal value of $f_{\rm c}$. No scaling collapse could be obtained here. }
\label{f:PertDep1}
\end{figure}

\begin{figure}[tbh]
{\setlength{\unitlength}{1cm}\begin{picture}(10,6)
\put(0,-0.5){ \includegraphics[width=\columnwidth]{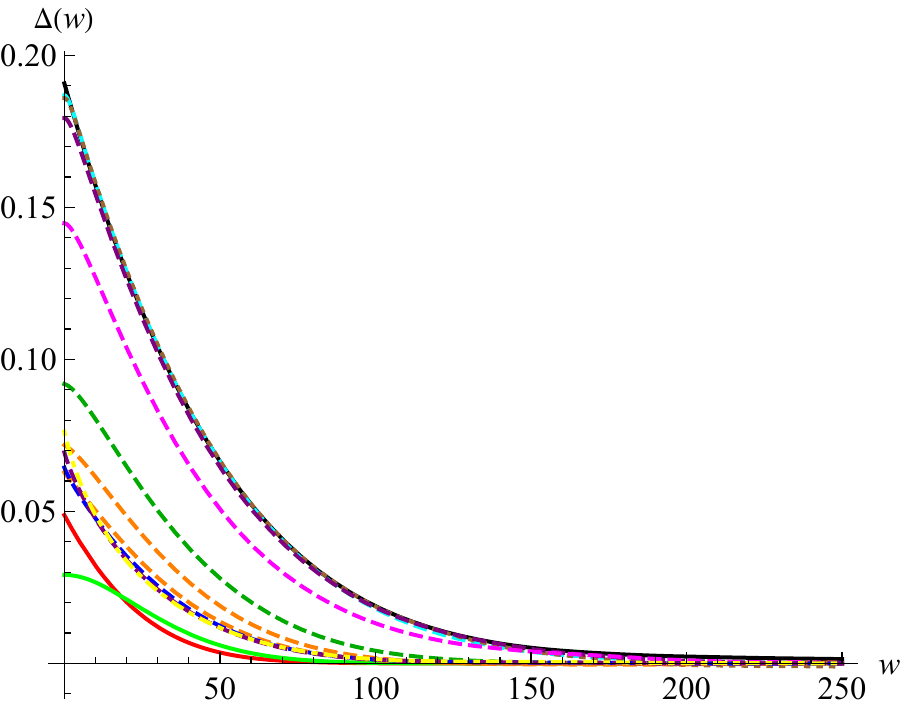}}
\put(2.5,1.5){ \includegraphics[width=0.6\columnwidth]{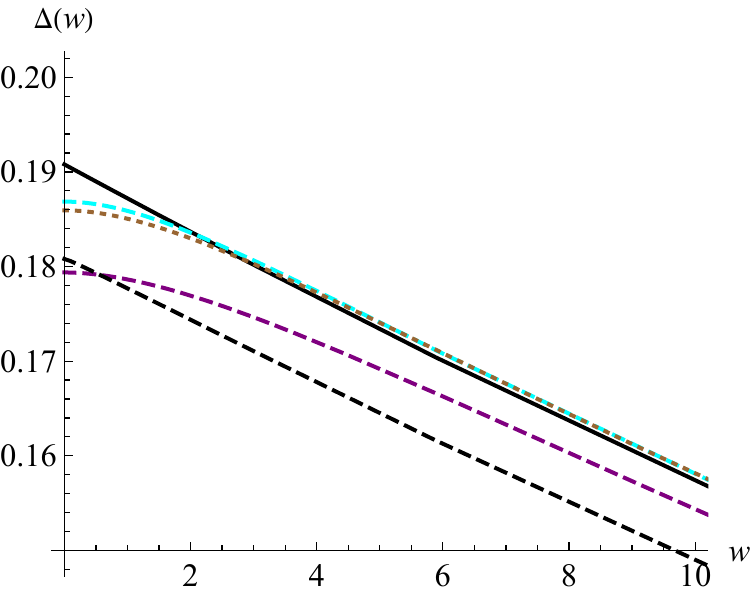}}
\end{picture} 
}  
\caption{$\Delta_{v, T}(w)$ (DNS) for $m^2 = 0.01$ at fixed $v= 0.1$ and varying $T$ compared to the depinning (black, top) and equilibrium (red, bottom, EM) and thermal rounding at $T = 5$. From top to bottom temperature increases $T \in [0, 0.001, 0.01, 0.1, 0.5 ,1 ,2, 3,4,5 ]$. At $T = 3$, the velocity boundary layer disappears due the formation of the thermal peak. Inset shows the small temperature effect on the boundary layer. It is little affected at small $T = 0.001$. Inset black dashed shows $T = 0.01, v = 10^{-3}$ and purple dashed $T = 0.01, v = 10^{-3}$ showing they are not related by velocity deconvolution.     }
\label{f:m2v1}
\end{figure}

In Figs.\ \ref{f:PertDep1} and \ref{f:m2v1} we show the whole crossover regime from depinning to equilibrium. The previous section studied the change in correlation length and area as an order parameter, but there is more we can say close to depinning. For this consider Fig. \ref{f:PertDep1}, at $v = 0.001$. In the inset we use the scaling relation \eqref{DeltaTScaling1} to collapse the curve for $T = 0.02$ onto the one for $T = 0$. In particular, this implies that the shape $\tilde{\Delta}(w)$ is not affected by temperature. When comparing experimental data to     theoretical predictions, this is important as scales are fixed using the correlation length. At larger $T$ this no longer holds true, and the shape changes. Another interesting feature can be identified at a larger driving velocity. Consider Fig. \ref{f:m2v1} for $v = 0.1$, where   rounding due to a finite driving velocity is clearly present. As we now know that when approaching equilibrium a thermal peak forms, one would expect some interplay between the velocity boundary layer and the thermal peak. Fig.\ \ref{f:m2v1} shows that this is indeed the case. For large $T >3$ an apparent cusp seems to re-emerge. Its nature, however, is very different from the cusps of the depinning and equilibrium fixed points. There it is related to the existence of shocks and avalanches. Here, it is an artefact of the combined effect of the velocity boundary layer and the thermal peak forming on top. This regime corresponds  to $\hat{T} = 0.32$, which is far in the crossover regime of Fig.\ \ref{f:ForceCollapse}. 

\section{Summary and discussion}
In this work we addressed the long-standing question of the full crossover between depinning and  equilibrium. Studying the force and the effective force correlator for a one-particle model,  we characterized the  phase diagram of finite velocity $v$ and finite temperature $T$. This may serve as a reference point for experiments and simulations in   dimensions $d>0$. We showed that the mean  force, divided by the mean force at depinning, is a robust    order parameter,  allowing one  to quantify where one is in between depinning and equilibrium, and what   one should expect for the force correlations.

Our results are directly applicable to the unzipping of a  DNA hairpin \cite{terBurgRissonePastoRitortWiese2022}. This experiment has all the ingredients studied here: It has a  finite temperature, it has random forces, and  it has a confining potential whose minimum is slowly increasing  at a driving velocity $v$,  allowing us to measure its force  correlations. Earlier analysis \cite{HuguetFornsRitort2009} has suggested this experiment to be close to equilibrium. Interestingly,  in this experiment   the stiffness of the trap, ($m^2$ in our notation) decreases when unzipping the DNA molecule,  
\be\label{m2ofw}
\frac1{m^2} = \frac1{m_0^2} + a n \ = \frac1{m_0^2} + a' w , 
\ee
where $n$ is the number of unzipped bases, itself proportional to the position of the confining potential $w$, starting with $w=0$ for the completely closed molecule. Reminding that $m$ sets the renormalization scale, we see that       the experiment    runs the renormalization group for   us! 
\begin{figure}[t]
{\setlength{\unitlength}{1cm}\begin{picture}(10,6)
\put(0,-0.5){ \includegraphics[width=\columnwidth]{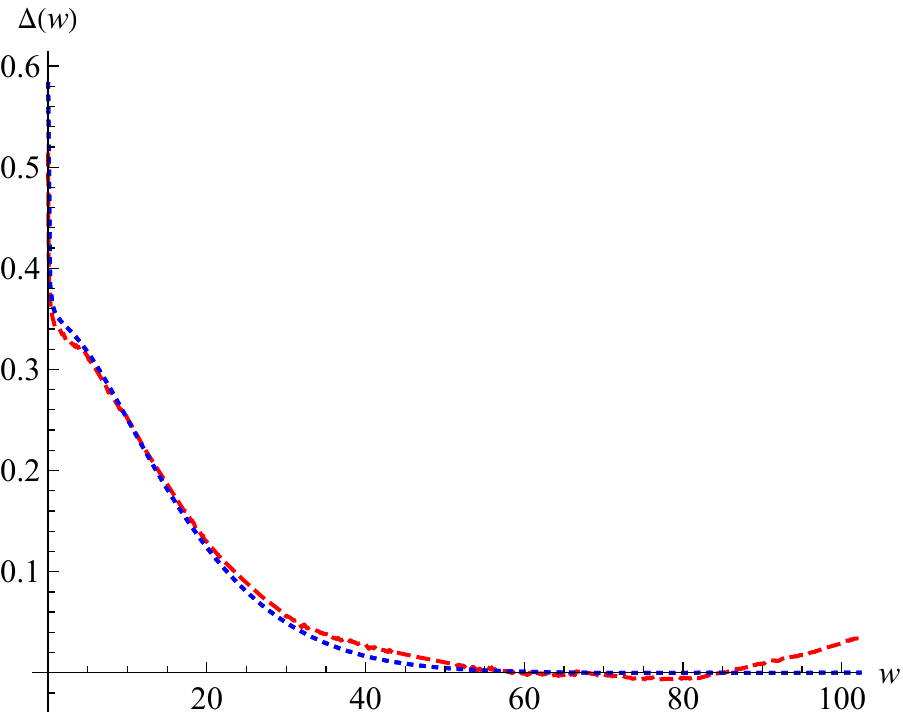}}
\put(2.3,1.3){ \includegraphics[width=0.7\columnwidth]{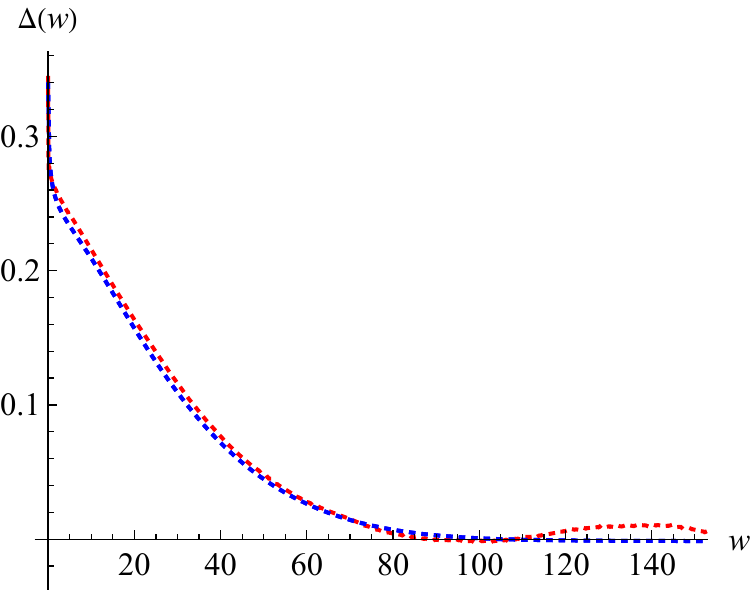}}
\put(4,4.3){Molecule almost unzipped}
\put(4,0.3){Beginning of   unzipping}
\end{picture} 
}  
\caption{Comparison of the experimental curve (red) to simulation (blue). For the main plot , simulation parameters are chosen to agree with the largest mass, i.e.\ small $w$ in \Eq{m2ofw}. In the inset we see results when the DNA molecule is almost unzipped, corresponding to a reduction in $m$ by a factor of  about $ 0.4$. Again  the  simulation    agrees     well with the experiment, and captures the shape of the boundary layer.}
\label{f:felix}
\end{figure} 
Figure \ref{f:felix} shows a comparison of experimental data for two different masses to numerical simulations. For the largest mass, simulation parameters are chosen such as to agree with experimental data. Using the ratio of masses in the experiment, this then predicts the simulation parameters for the smaller mass. We see that simulation and experiment agree well.  We report more on this experiment in \cite{terBurgRissonePastoRitortWiese2022}. 

We hope that this work serves as a reference  where both the driving velocity $v$ and temperature $T$ are non-vanishing, and it is a priori not clear where in the phase diagram one is sitting.  Looking at the measured critical force divided by its value at depinning allows one to identify where in the phase diagram an experiment is located. One can then asses and quantify
 all the features discussed here: Thermal rounding,  the thermal peak and its broadening as a function of    $m^2$, as well as the scaling length $\rho_m$ in the $w$ direction. This should be useful in order to bring some order into these many-parameter systems.

\acknowledgements
We thank A.~Kolton and A.~Rosso for sharing their experience, and P.~Rissone, M.~Rico-Pasto and F.\ Ritort for the experimental collaboration for DNA unzipping.

%
%
\appendix

\section{Numerical implementations}
\label{s:Numerical implementations}

The number of samples is denoted by $N$. 
In this work we use two numerical implementations:
\begin{enumerate}
\item[(i)] {\em Direct numerical simulation (DNS)}.
To solve the coupled set of differential equations  \eq{EOM}-\eq{FEOM}
we  use a space discretization $\delta u =10^{-2}$ to first obtain  the random forces $F(u)$ for $u=n \delta u$, $n\in \mathbb N$. $F(u)$  is then linearly interpolated  between these  points.  We finally  solve \Eq{EOM} with the Euler method, using a time-discretization of $\delta t = 10^{-3}$.  
\item[(ii)] {\em Exact minimisation (EM)}. 
In the statics at temperature $T=0$, the relevant quantities are computed using minimisation of the energy in \Eq{EnergyMinimisation}. For a given disorder realisation $V(u)$, the minimum of the potential as a function of $w$ is 
\be
 \hat{V}(w)  =    \text{ min}_u \biggl[ V(u) + \frac{m^2}{2}(u - w)^2 \biggl],  \label{VeffFormula} 
\ee
At finite temperature, this is replaced by 
\be
 \qquad \hat{V}(w) 
 =   V(w) - T  \ln{\biggl( \langle e^{- \frac{V(u){-} V(w)}{T} - \frac{m^2}{2T}(u {-}w )^2} \rangle_u \biggl)} .    \!\!\! 
\ee
Using potential differences allows to better  restrict the necessary range in $u-w$. 
For RF disorder, as for OU forces, the (microscopic) potential is obtained by integrating the random forces,  
\be 
V(u) -V(w) = - \int_w^u F(u^\prime) \rmd u^\prime. 
\ee
The effective force $\hat{F}(w) = - \partial_w \hat{V}  $ then becomes
\begin{equation}
\quad \quad  \hat{F}(w) =  m^2  \frac{ \langle e^{- \frac{V(u) - V(w)}{T} - \frac{m^2}{2T}(u -w )^2} (u -w)\rangle_u    }{\langle e^{- \frac{V(u) - V(w)}{T} - \frac{m^2}{2T}(u -w )^2} \rangle_u  }.  \label{FeffFormula} 
\end{equation}
\end{enumerate}

\section{Boundary layer}
\label{A:Boundary layer}

At finite temperature, the unrescaled 1-loop FRG equation   acquires  an additional term,  
\bea
-m \partial_m  \Delta (w) &=&  
  -   \frac{1}{2}  \partial_w^2\bigl[ {\Delta}(w) {-} {\Delta}(0) \bigl] ^2   
+ \tilde{T}_m \Delta_m^{\prime \prime }(w)  \dots \qquad 
\label{1loopFRGTa} \\
\tilde{T}_m & : = &     {2T m^\theta}     \int_k \frac{ 1}{k^2+m^2}\biggl|_{m = 1} .
 \label{B2}
\eea
(In dimension $d=0$, the integral simplifies to $1/m^2$).
The fixed-point equation for the rescaled dimensionless disorder $\tilde \Delta(w) := m^{\epsilon-2 \zeta }\Delta(w m^\zeta)$ then takes the form
\bea \label{1loopFRGTb}
 -m \partial_m \tilde \Delta ({ w}) &=&    (\epsilon  {-} 2 \zeta) \tilde{\Delta}({ w}) + \zeta { w} \tilde{\Delta}^\prime({ w})  \\
  && -   \frac{1}{2}  \partial_{ w}^2\bigl[ \tilde{\Delta}({ w}) {-} \tilde{\Delta}(0) \bigl] ^2   
+ \tilde{T}_m \tilde{\Delta}^{\prime \prime }({ w})  \dots \nn
\eea
What is remarkable about \Eq{1loopFRGTa} is that the RG flow conserves the integral $\int_{w>0} \Delta(w)$, both at vanishing temperature $\tilde T_m=0$ and at  $\tilde T_m>0$. The reason is that the r.h.s.\ of \Eq{1loopFRGTa} is a total derivative. 

\begin{figure}[t]
 \includegraphics[width=\columnwidth]{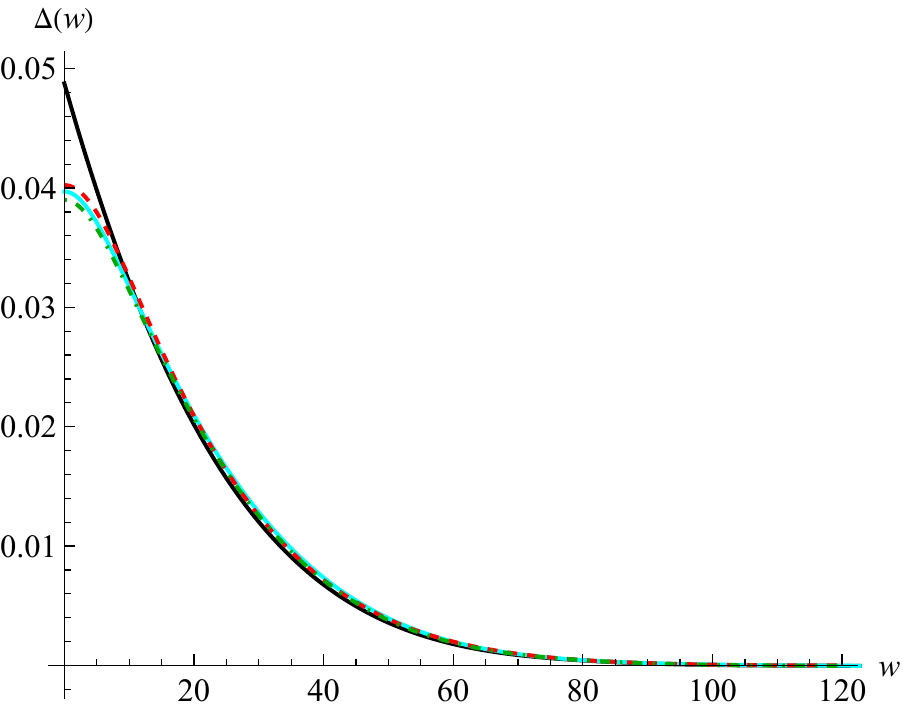}  
\caption{Comparison of the boundary layer (dark-green dot-dashed) to the diffusion kernel (red-dashed), experimental data at $T = 1, m^2 =0.01$ in cyan and the zero temperature fixed point in black. }
\label{f:DiffKernel}
\end{figure} 
For the random-field solution $\zeta=\epsilon/3$ in equilibrium, relevant for us, this also holds for the rescaled  \Eq{1loopFRGTb}. 

The  finite-temperature solution in the standard  boundary-layer form  is \cite{Wiese2021}
\begin{align} 
\ & \Delta_{T}(w)  \approx      \mathcal{A}_{T}  \Delta(\sqrt{w^2 + t^2})  \label{D3}\\
t \ = &   \frac{6   T m^2 }{\epsilon |\Delta^\prime(0^+)|} \quad  \Leftrightarrow 
\quad 
\frac{t}{\rho_m} = \frac{6 T m^2}{\epsilon \Delta(0)}.  \qquad \quad  \label{tpred}
\end{align}
As the flow preserves the area, it is important to fix $\mathcal{A}_{T}$, s.t.\ the integrals on both sides coincide. 
This adds a non-trivial   change in normalization which cannot be given in closed form. 
Another problem of the boundary layer is that given $\Delta_{T}(w)$, one can  reconstruct $\Delta(w)$ only for $w \ge t$. 
Since the boundary layer is phenomenological and not exact, we   propose a different approximation: namely, to obtain the finite-$t$ solution   by convoluting the zero-temperature solution with an appropriately chosen diffusion kernel,   
\bea\label{DiffKernelsupp-app} 
\Delta_{T}(w)  &=&    \int_{-\infty}^{\infty} \rmd u\,  \Delta(u) G(u-w,\tau), \\
G(u,\tau)  &=& \frac{1}{ \sqrt{4\pi \tau}} \rme^{-\frac{u^2}{4 \tau }}.  
\eea
A nice property of the convolution   in \Eq{DiffKernelsupp-app} is that   by construction it is area preserving, thus no additional normalization $\mathcal{A}_{T}$ is necessary. While using the diffusion kernel is natural, given that \Eq{1loopFRGTb} is the diffusion equation in absence of non-linear terms, 
what remains to be done is to fix the   ``diffusion time'' $\tau$. Given the properties of the diffusion kernel, this can analytically be done for 
\be
\Delta(w) = \ca C  \rme^{- w/\rho_m  - b (w/\rho_m)^2}.
\ee
Demanding that $\Delta_{T}''(0)/\Delta_{T}(0)$ agree yields
\be
\tau = \frac{t^2}{\pi} - \frac{2(\pi - 2)t^3}{\rho_m \pi^2} +  \mathcal{O}(t^4).
\ee
The leading-order term only depends on $t$, while the subleading one contains $\rho_m$. Higher-order terms   depend on $b$.

Fig.~\ref{f:DiffKernel}   shows a comparison of   numerics for $m^2 = 0.01$ at  $T = 0$ (black) and $T = 1$ (cyan) to the boundary-layer approximation  \eq{D3} (dark-green dot-dashed) and the diffusion kernel \eqref{DiffKernelsupp-app} (red dashed), with   $t$   from \Eq{tpred}. Both approximations seem to work   well.  

\section{Exact relation between microscopics and macroscopics}
\label{app3}

The FRG equation \eq{1loopFRGTa} predicts that the integral $\int \Delta(w) {\rm d}w $ remains unrenormalised. 
Therefore the integral over the microscopic disorder $\Delta_0(w)$ equals the integral over the renormalized disorder $\Delta(w)$, which   we can rewrite through its scaling form \eq{5}  as
\bea
 \int_0^\infty \rmd w \, \Delta_{0}(w) &  \equiv &   \int_0^\infty \rmd w \, \Delta (w) \label{noRenorm} \nn \\
&= &  \int_0^\infty \rmd w\, m^4 \rho_m ^2 \tilde \Delta(w/\rho_m ) \nn  \\ 
 & = &\text{ }  m^4 \rho_m^3 \int_0^\infty \rmd w\, \tilde \Delta(w) . 
\eea
Since $\rho_m \sim m^{-\zeta}$,   the combination $m^4 \rho_m^3$ is independent of $m$  for RF disorder which has $\zeta=4/3$. (Note that this also works in dimension $d>0$,  with $m^4$ in \Eq{noRenorm} replaced by $m^\epsilon$, $\epsilon=4-d$, and $\zeta=\epsilon/3$.)
Solving for $\rho_m$ we find  
\be
\rho_m=  \left[ \frac{  \int_{w>0}  \Delta_{0}(w) }{m^{4} \int_{w>0} \tilde \Delta(w) }\right]^{1/3} .
\ee
For equilibrium RF disorder in $d=0$ (see section \ref{s:TempEquilibrium}),  $\int_{w>0} \tilde{ \Delta}(w)=0.252$, and this reduces to
\be
\rho_m =   \left[ \frac {3.97}  {m^4} \int_{w>0}  \Delta_{0}(w)   \right]^{1/3}. \label{RhoPred}
\ee
Eq.~\eqref{noRenorm} has been  verified experimentally in Ref.~\cite{terBurgRissonePastoRitortWiese2022}. Here we perform a numerical test. For the simulations of section \ref{f:sinai} microscopic forces are taken constant on an interval of size one, with variance 1. As a consequence, the microscopic disorder has integral $\int_{w>0} \Delta_0(w)   = 1/2$. Numerical simulations of \Eq{EnergyMinimisation} confirm that this is preserved under RG:   $\int_{w>0} \Delta(w)  = 0.496 $ for $m^2 = 10^{-2}$, $\int_{w>0} \Delta(w)   w = 0.484 $ for $m^2 = 10^{-3}$ and $\int_{w>0} \Delta(w)   = 0.525 $ for $m^2 = 10^{-4}$.  Using Eq.~\eqref{RhoPred} this gives a prediction for the scale $\rho_m$. This confirms for a single particle   that, if the microscopic disorder is known, there are no unknown scales. Both $\rho_m$ as well as $\Delta(0)$ are predicted by the microscopic disorder.

\ifx\doi\undefined
\providecommand{\doi}[2]{\href{http://dx.doi.org/#1}{#2}}
\else
\renewcommand{\doi}[2]{\href{http://dx.doi.org/#1}{#2}}
\fi
\providecommand{\link}[2]{\href{#1}{#2}}
\providecommand{\arxiv}[1]{\href{http://arxiv.org/abs/#1}{#1}}
\providecommand{\hal}[1]{\href{https://hal.archives-ouvertes.fr/hal-#1}{hal-#1}}
\providecommand{\mrnumber}[1]{\href{https://mathscinet.ams.org/mathscinet/search/publdoc.html?pg1=MR&s1=#1&loc=fromreflist}{MR#1}}

\tableofcontents

\end{document}